\documentclass[12pt,thmsa]{article}
\usepackage{sw20lart}

\input{tcilatex}
\begin{document}

\begin{center}
\smallskip

{\Large Universal construction of quantum computational networks in
superconducting Josephson junctions} \smallskip \ \\[0pt]
\ 

Xijia Miao \\[0pt]
Laboratory of Magnetic Resonance and Atomic and Molecular Physics, Wuhan
Institute of Physics and Mathematics, The Chinese Academy of Sciences, Wuhan
430071, People$^{^{\prime }}$s Republic of China\ 

E-mail: miao@nmr.whcnc.ac.cn \ \\[0pt]
\end{center}

\ \newline
\textbf{Abstract }\newline

Any quantum computational network can be constructed with a sequence of the
two-qubit diagonal quantum gates and one-qubit gates in two-state quantum
systems. The universal construction of these quantum gates in the quantum
systems and of the quantum computational networks with these gates may be
achieved with the help of the operator algebra structure of Hamiltonians of
the systems and the properties of the multiple-quantum operator algebra
subspaces of the Liouville operator space and the specific properties of the
quantum algorithm corresponding to the quantum network. As an example, the
two-qubit diagonal gates are exactly prepared in detail in superconducting
Josephson junctions.\newline
\newline

\smallskip

\smallskip

Quantum computation becomes increasingly attractive largely due to the
reason that it can be much more powerful than the classical counterpart and
has an extensively potential application.$^{1-4}$ Quantum computational
networks are usually composed of a sequence of quantum gates that could be
built up with one- and two-body interactions of quantum systems.$^{5-7}$ The
practical quantum gates should satisfy several requirements. For example,
they should be simple so that any quantum computational network can be built
easily out of these gates $^{5,\text{ }8}$ and could be conveniently
prepared experimentally in those accessible quantum systems. Owing to
inevitable decoherence and dephase effects in practical quantum systems the
quantum gates should be chosen suitably in order to minimize thses effects
on the quantum computation built up with these gates. The practical quantum
gates may include the three-qubit Toffoli gate,$^{9}$ the Deutsch
there-qubit universal gate,$^{5}$ the two-qubit universal gate,$^{10,11}$
the two-qubit XOR gate $^{12}$ and the two-qubit diagonal gate,$^{13}$ etc.
In recent years, a variety of two-state quantum systems have been
investigated experimentally to construct the quantum gates. These systems
include trapped ions,$^{14}$ cavity quantum electrondynamics,$^{15}$ nuclear
spin systems in molecules,$^{16,17}$ and the superconducting
Josephson-junction arrays.$^{18-21}$ Quantum computation obeys quantum
physical laws and time evolution of a quantum system during quantum
computing is described by the Schr$\ddot{o}$dinger equation, $^{22}$ while a
quantum algorithm whose quantum network usually consists of a sequence of
unitary transformations places constraints on the time evolution and hence
constraints the form of the effective Hamiltonian of the quantum system in
the quantum computing.$^{13}$ A quantum algorithm is usually more dependent
on mathematical and quantum physical principles and largely independent of a
general quantum system, but the quantum-gate construction is always
performed in a practical quantum system. Therefore, in real quantum
computation it is important to consider the specific properties of both
quantum systems and quantum algorithms for the unified and systematical
construction of the quantum gates in a variety of accessible quantum systems
and of any quantum networks with these gates. However, this point is often
neglected and is really not paid much attention to. In fact, such a
universal construction may be achieved simply with the help of the operator
algebra structure of the Hamiltonian and the properties of the
multiple-quantum operator algebra subspaces of the Liouville operator space
of a quantum system, and the specific properties of a given quantum
algorithm.$^{23}$ In this communication it is shown how to build exactly the
two-qubit diagonal quantum gates in superconducting Josephson junction
arrays $^{18-21}$ by making use of the operator algebra structure of
Hamiltonian and the properties of the multiple-quantum operator algebra
subspaces of the quantum system. This general principle allows one to apply
to other coupled two-state multiparticle quantum systems.

The quantum computational network of a given quantum algorithm usually
consists of a sequence of quantum circuit units.$^{5}$ Each such unit (e.g.,
the $k$th unit) is described by a unitary operator that may be written as an
exponential form: $U_{k}(t_{k})=\exp (-iH_{k}t_{k})$, where $H_{k}$ is the
time-independent effective Hamiltonian associated with the $k$th circuit
unit. To implement the quantum circuit units in a quantum system one
possible strategy is that the unitary exponential operator $U_{k}(t_{k})$ is
first decomposed completely into an ordered product of a series of
elementary building blocks $G_{m}(\lambda _{m})$, i.e., the one-qubit gates
and the two-qubit diagonal quantum gates $^{13}$

$\qquad \qquad \qquad \qquad \qquad U(t)=\stackunder{m}{\prod }G_{m}(\lambda
_{m})\qquad \qquad \qquad \ \ \qquad \qquad \quad \ \ \ (A)$ \newline
The two-qubit diagonal gate $G_{kl}(\lambda _{kl})$ is defined in an
operator form by

$\qquad \qquad \qquad \qquad G_{kl}(\lambda _{kl})=\exp (-i\lambda
_{kl}2I_{kz}I_{lz})\qquad \qquad \qquad \quad \ \qquad \ (B1)$ \newline
or in the unitary representation in the conventional computational basis by

$\qquad G_{kl}(\lambda _{kl})=Diag[e^{-i\frac{1}{2}\lambda _{kl}},\ e^{i%
\frac{1}{2}\lambda _{kl}},\ e^{i\frac{1}{2}\lambda _{kl}},\ e^{-i\frac{1}{2}%
\lambda _{kl}}]\qquad \qquad \quad \ \ \qquad (B2)$ \newline
where the operators $I_{pz}=\frac{1}{2}\sigma _{pz}\ (p=k,l),\ \sigma $ is
Pauli$^{^{\prime }}$s operator and the two-qubit diagonal gate acts on only
the $k$th and $l$th qubits in the quantum system simultaneously. The quantum
computation then can be carried out by implemeting the decomposed unitary
operations $U_{k}(t_{k})$ of Eq.(A) in a feasible quantum system. The
decomposition of Eq.(A) may be achieved in an exact and unified form and may
be simplified greatly with the help of the operator algebra structure of the
Hamiltonian $H_{k}$ and the properties of the Liouville operator space and
its multiple-quantum operator subspaces.$^{23}$ If the Hamiltonian $H_{k}$
is a member of the \textit{lo}ngitudinal \textit{m}agnetization and \textit{s%
}pin \textit{o}rder (\textit{LOMSO}) operator subspace the decomposition of
Eq.(A) is simple due to the fact that each pair of the base operators of the
operator subspace commute. When the Hamiltonian $H_{k}$ is a zero-quantum
operator, the decomposition is carried out in the zero-quantum operator
subspace, while $H_{k}$ is an \textit{ev}en-\textit{o}rder \textit{m}ultiple-%
\textit{q}uantum (\textit{EVOMQ}) operator the decomposition may be achieved
in the \textit{EVOMQ} operator subspace. For example, all the diagonal phase
quantum gates can be constructed in the \textit{LOMSO} operator subspace,$%
^{3-5,13}$ while an arbitrary SWAP operation $^{5}$ between any pair of
quantum bits may be built up in the zero-quantum operator subspace.

The preparation for the elementary building blocks, that is, the two-qubit
diagonal gates in a variety of two-state coupled multiparticle quantum
systems may also be simplified with the help of the operator algebra
structure of the Hamiltonians of the quantum systems, and the properties of
the Liouville operator space and its multiple-quantum operator subspaces.
For example, for a weak coupled N-spin (I=1/2) system $^{24,25}$ its
Hamiltonian is a memeber of the \textit{LOMSO} operator subspace. The
two-qubit diagonal gate then can be achieved with the aid of the properties
of the \textit{LOMSO} subspace that any pair of base operators of the
subspace commute, and may be prepared exactly by using a sequence of spin
echo units with selective radiofrequency pulses in the system.$^{13}$
Likewise, if the Hamiltonian is an element of the zero-quantum operator
subspace the elementary building blocks may be prepared by utilizing the
properties of the zero-quantum operator subspace. The Hamiltonians of the
superconducting Josephson junction system may consist of the
multiple-quantum operators (see below). Then the properties of the
multiple-quantum operator subspaces, e.g., the \textit{EVOMQ} subspace will
be helpful for the exact preparation of the two-qubit diagonal gate in the
system.

It has been suggested in recent years that superconducting Cooper-pair boxes
with Josephson junctions could be used as qubits.$^{18-21}$ Very recently,
the coherent time evolution of the two charge states in a single-Cooper-pair
box is observed experimentally.$^{21}$ This demonstrates for the first time
a practical solid-state qubit for quantum computation. It is discussed in
detail below how to prepare the two-qubit diagonal gate in the
low-capacitance SQUID-controlled Josephson junction arrays.$^{18,20}$
Obviously, any one-qubit gate is easily implemented in the system and will
not be further discussed. Now consider a system of two coupled
SQUID-controlled Josephson junctions, without loss of generality. The
Hamiltonian for the system can be written generally in a spin (I=1/2)
language $^{18,20}$

$H_{T}=E_{ch}^{k}(V_{x_{k}})\sigma _{z}^{k}+E_{ch}^{l}(V_{x_{l}})\sigma
_{z}^{l}-\frac{1}{2}E_{J}^{k}(\Phi _{x_{k}})\sigma _{x}^{k}-\frac{1}{2}%
E_{J}^{l}(\Phi _{x_{l}})\sigma _{x}^{l}$

$\qquad \qquad -[E_{J}^{k}(\Phi _{x_{k}})E_{J}^{l}(\Phi
_{x_{l}})/E_{L})\sigma _{y}^{k}\sigma _{y}^{l}\qquad \qquad \qquad \qquad
\qquad \qquad \qquad (1)$ \newline
where $\sigma $ is Pauli$^{^{\prime }}$s operator. By a simple unitary
transformation: $H=U_{y}^{+}H_{T}U_{y}$ where$\ U_{y}=\exp (-i\phi
_{k}I_{ky})\exp (-i\phi _{l}I_{ly})$, the Hamiltonian (1) can reduce to a
simpler form

$\qquad \qquad \ \ H=\Omega _{k}I_{kz}+\Omega _{l}I_{lz}+\pi
J_{kl}2I_{ky}I_{ly}\qquad \qquad \qquad \qquad \qquad \qquad (2)$ \newline
where the operators $I_{i\mu }=\frac{1}{2}\sigma _{\mu }^{i}\ ,\ \tan \phi
_{i}=2E_{ch}^{i}(V_{x_{i}})/E_{J}^{i}(\Phi _{x_{i}}),\ $\newline
$\Omega _{i}=-2E_{ch}^{i}(V_{x_{i}})\sin \phi _{i}-E_{J}^{i}(\Phi
_{x_{i}})\cos \phi _{i}\ (i=k,l;\ \mu =x,y,z),\ $and $\pi
J_{kl}=-2E_{J}^{k}(V_{x_{k}})E_{J}^{l}(V_{x_{l}})/E_{L}$. By employing the
propagator $U(t)=\exp (-iHt)$ corresponding to the Hamiltonian (2) one can
prepared the two-qubit diagonal gates based on the Baker-Campbell-Hausdoff
formula, correct to the third order approximation,$^{13}$

$\ G_{kl}(\lambda _{kl})=\exp (-i\frac{\pi }{2}F_{x})\exp (-iHt/2)\exp
(-i\pi F_{z})\exp (-iHt)$

\qquad $\qquad \qquad \times \exp (i\pi F_{z})\exp (-iHt/2)\exp (i\frac{\pi 
}{2}F_{x})$

$\qquad \qquad =\exp (-i2\pi J_{kl}2I_{kz}I_{lz}t)+O(t^{3})\qquad \qquad
\qquad \qquad \qquad \ \ \qquad (3)$ \newline
where the operators $F_{\mu }=I_{k\mu }+I_{l\mu }\ (\mu =x,y,z)$ and the
unitary operations $\exp (\pm i\alpha F_{\mu })$ are single-qubit operations.

Actually, the two-qubit diagonal gates of Eq.(B) can be prepared exactly
with the help of the operator algebra structure of the Hamiltonian of the
superconducting Josephson junction arrays and the properties of the
multiple-quantum operator subspaces. By a simple transformation: $H_{e}=\exp
(i\frac{\pi }{2}F_{y})H\exp (-i\frac{\pi }{2}F_{y})$, the Hamiltonian (2)
can be converted into a member of the \textit{EVOMQ} operator subspace

$H_{e}=\Omega _{k}I_{kz}+\Omega _{l}I_{lz}+\pi
J_{kl}(I_{kx}I_{lx}+I_{ky}I_{ly})+\pi
J_{kl}(-I_{kx}I_{lx}+I_{ky}I_{ly})\qquad (4)$ \newline
where $I_{kz}\ $and $I_{lz}$ are the longitudinal magnetization operators, $%
(I_{kx}I_{lx}+I_{ky}I_{ly})$ and $(-I_{kx}I_{lx}+I_{ky}I_{ly})$ the zero-
and double-quantum operators, respectively, $^{24,25}$ indicating that the
operator $H_{e}$ is an even-order multiple-quantum operator. Then by
utilizing the properties of the \textit{EVOMQ} operator subspace one can
diagonalize the operator H$_{e}$ in the operator subspace. $^{23}$ The
result turns out to be as follows

$\qquad \qquad \qquad \qquad \widetilde{H}=V^{+}H_{e}V=\Omega _{k}^{^{\prime
}}I_{kz}+\Omega _{l}^{^{\prime }}I_{lz}\qquad \qquad \qquad \qquad \qquad
(5) $ \newline
where the unitary operator $V$ is given by

$\qquad \qquad \qquad \quad V=\exp (-i\alpha Q_{0})\exp (-i\beta
Q_{2})\qquad \qquad \qquad \ \ \qquad \qquad (6)$ \newline
with the zero-quantum operator $Q_{0}$ and the double-quantum operator $%
Q_{2} $: $Q_{0}=2(I_{kx}I_{ly}-I_{ky}I_{lx})$ and $%
Q_{2}=2(I_{kx}I_{ly}+I_{ky}I_{lx})$.

For simplification, the ratio $\gamma $ between the parameters $\Omega
_{k}^{^{\prime }}$ and $\Omega _{l}^{^{\prime }}$ in Eq.(5) is set as $%
\gamma =(2m+1)/(2m)\ (m=\pm 1,\pm 2,...),$ then

$\qquad \qquad \qquad \qquad \qquad \qquad \Omega _{k}^{^{\prime }}=\gamma
\Omega _{l}^{^{\prime }},\qquad \qquad \qquad \qquad \qquad \qquad \qquad
(7) $ \newline
and in the system of the two coupled SQUID-controlled Josephson junctions
the evolutional time t is chosen as $t_{p}$ so that

$\qquad \qquad \qquad \qquad \qquad \Omega _{k}^{^{\prime }}t_{p}=(2m+1)\pi
\qquad \qquad \qquad \qquad \qquad \qquad \ (8a)$ \newline
and hence

$\qquad \qquad \qquad \qquad \qquad \Omega _{l}^{^{\prime }}t_{p}=2m\pi
.\qquad \qquad \qquad \qquad \qquad \qquad \qquad \ \ (8b)$\newline
Note that $\Omega _{k}^{^{\prime }}$ and $\Omega _{l}^{^{\prime }}$ are
dependent on the parameters $\Omega _{k}$, $\Omega _{l}$, and $J_{kl}$. If $%
\Omega _{i}\ (i=k,l)$ and $J_{kl}$ are adjusted suitably, this could be
achieved by tuning the external flux $\Phi _{x_{i}}$ and the inductance L in
the superconducting Josephson junction circuit,$^{18,20}$ equation (7) can
be met for a given ratio $\gamma $. By Eqs.(5)-(8) the propagator
corresponding to the Hamiltonian $H_{e}$ of Eq.(4) can take a simpler form

$\qquad \qquad \qquad \qquad \exp (-iH_{e}t_{p})=V^{2}\exp (-i\pi I_{kz}).\
\ \ \quad \qquad \quad \qquad \qquad (9)$ \newline
In order to prepare the two-qubit diagonal gate an echo sequence is
constructed as follows

$\qquad \qquad P_{1}=\exp (-iH_{e}t_{p})\exp (-i\pi I_{kx})\exp
(-iH_{e}t_{p}).\qquad \qquad \qquad \ \ (10)$ \newline
With the help of Eqs.(6) and (9) this operation is simplified

$\qquad \qquad P_{1}=\exp [-i8(-\alpha +\beta )I_{ky}I_{lx}]\exp (-i\pi
I_{kx}).\qquad \qquad \qquad \qquad (11)$\newline
Then one obtains the two-qubit diagonal gate from Eqs.(10) and (11)

$G_{kl}(\lambda _{kl})=T_{1}\exp (-iHt_{p})\exp (i\pi I_{kz})\exp
(-iHt_{p})T_{1}^{+}\exp (i\pi I_{kx})\qquad \ \ \ (12)$ \newline
where $\gamma =(2m+1)/(2m)$, the evolutional time $t_{p}=(2m+1)\pi /\Omega
_{k}^{^{\prime }}$, the parameter $\lambda _{kl}=4(-\alpha +\beta ),$ and
the unitary operation $T_{1}$ represents the single-qubit composite rotation
operation:

$\qquad \qquad \qquad \ \ T_{1}=\exp (-i\frac{\pi }{2}I_{kz})\exp (i\frac{%
\pi }{2}I_{kx}).\qquad \qquad \qquad \ \ \ \qquad \qquad (13)$ \newline
To determine the parameters $\Omega _{k}^{^{\prime }}$, $\Omega
_{l}^{^{\prime }}$ and $\alpha ,\ \beta $ one needs to expand the
transformation $V^{+}H_{e}V$ of Eq.(5) by using the rotation transformation
between any two product operators.$^{25}$ The equations to determine these
parameters are therefore obtained explicitly

$\Omega _{k}\sin \alpha _{1}\cos \alpha _{2}+\Omega _{l}\cos \alpha _{1}\sin
\alpha _{2}-\pi J_{kl}\sin \alpha _{1}\sin \alpha _{2}=0\qquad \qquad \
\quad (14a)$

$\Omega _{k}\cos \alpha _{1}\sin \alpha _{2}+\Omega _{l}\sin \alpha _{1}\cos
\alpha _{2}+\pi J_{kl}\cos \alpha _{1}\cos \alpha _{2}=0\qquad \quad \qquad
(14b)$\newline
and

$\Omega _{k}^{^{\prime }}=\Omega _{k}\cos \alpha _{1}\cos \alpha _{2}-\Omega
_{l}\sin \alpha _{1}\sin \alpha _{2}-\pi J_{kl}\cos \alpha _{1}\sin \alpha
_{2}\qquad \qquad \ (15a)$

$\Omega _{l}^{^{\prime }}=\Omega _{l}\cos \alpha _{1}\cos \alpha _{2}-\Omega
_{k}\sin \alpha _{1}\sin \alpha _{2}-\pi J_{kl}\sin \alpha _{1}\cos \alpha
_{2}\qquad \qquad \ \ (15b)$ \newline
where $\alpha _{1}=-\alpha +\beta $ and $\alpha _{2}=\alpha +\beta $. It is
easy to find two independent solutions $\alpha _{2}=\alpha _{2}^{\pm }$ to
Eqs.(14) that are explicitly expressed as

$\qquad \qquad \qquad \tan \alpha _{2}^{\pm }=\frac{1}{2}(-\delta \pm \sqrt{%
\delta ^{2}+4})\qquad \qquad \qquad \qquad \qquad \qquad (16a)$\newline
and the corresponding $\alpha _{1}=\alpha _{1}^{\pm }$ that are determined by

$\qquad \qquad \tan \alpha _{1}^{\pm }=-\dfrac{\pi J_{kl}}{\Omega _{l}}-%
\dfrac{1}{2}\dfrac{\Omega _{k}}{\Omega _{l}}(-\delta \pm \sqrt{\delta ^{2}+4}%
)\qquad \qquad \qquad \quad \qquad (16b)$\newline
where $\delta =[\Omega _{l}^{2}-\Omega _{k}^{2}+(\pi J_{kl})^{2}]/(\pi
J_{kl}\Omega _{k})$. Note that the propagator $exp(-iH_{e}t_{p})$ is the
same for a given time $t_{p}$ for the two solutions $(\alpha _{1}^{+},\
\alpha _{2}^{+})$ and $(\alpha _{1}^{-},\ \alpha _{2}^{-})$. Without lossing
generality, here the solution is taken as $(\alpha _{1}^{+},\ \alpha
_{2}^{+})$ and the parameters $\Omega _{k}^{^{\prime }}$ and $\Omega
_{l}^{^{\prime }}$ are calculated by inserting the solution into Eqs.(15).
Consequently, one can explicitly obtain equation (7) that constrains the
possible values of the physical parameters $\Omega _{k}$, $\Omega _{l}$, and 
$J_{kl}$ for a given $\gamma =\gamma ^{+}$ (it can prove that $\gamma \neq
0,\ \pm 1$). In theoretical design the parameters $\lambda _{kl}=4\alpha
_{1}^{+}$ and $\gamma ^{+}=(2m+1)/(2m)$ are given in advance. Then equations
(7) and (16b) can be simplified and recast a one-variable $(\mu )$ quadratic
(the parameters $\mu =\Omega _{l}/\Omega _{k}$ and $\nu =\pi J_{kl}/\Omega
_{k}$ below):

$[p-\tan (\frac{1}{4}\lambda _{kl})]^{2}\mu ^{2}-\gamma ^{+}\{1+2p[p-\tan (%
\frac{1}{4}\lambda _{kl})]+\tan ^{2}(\frac{1}{4}\lambda _{kl})\}\mu $

$\qquad \qquad \qquad \qquad \qquad \qquad \qquad +(p\gamma
^{+})^{2}+1=0\qquad \qquad \qquad \qquad (17a)$\newline
with

$\qquad p=\dfrac{\tan (\frac{1}{4}\lambda _{kl})[1+\tan ^{2}(\frac{1}{4}%
\lambda _{kl})]}{2[1+\tan (\frac{1}{2}\lambda _{kl})\tan (\frac{1}{4}\lambda
_{kl})]}$\newline
After the parameter $\mu $ is determined from Eq.(17a) another parameter $%
\nu $ can be obtained by

$\qquad \qquad \qquad \qquad \qquad \nu =p(\gamma ^{+}-\mu )\qquad \qquad
\qquad \qquad \qquad \qquad \quad \ (17b)$ \newline
In Eqs.(17) it is required that

$\mu \gamma ^{+}[1+\tan ^{2}(\frac{1}{4}\lambda _{kl})]>1$ \newline
and the solution $(\alpha _{1}^{+},\ \alpha _{2}^{+})$ also requires that

$\dfrac{\mu ^{2}+\nu ^{2}-1}{\nu }>2\mu \tan (\frac{1}{4}\lambda _{kl})+2\nu
.$\newline
Given the parameters $\lambda _{kl}$ and $\gamma ^{+}$, the parameters $\mu $
and $\nu $ can be determined by Eqs.(17) and consequently, the physical
parameters $\Omega _{k}$, $\Omega _{l}$, and $J_{kl}$ in the superconducting
Josephson junctions are set up properly, while the parameter $\Omega
_{k}^{^{\prime }}$ is calculated through Eq.(15a) by using these determined
parameters. The evolutional time $t_{p}=\left| (2m+1)\pi /\Omega
_{k}^{^{\prime }}\right| $ is therefore obtained exactly.

In practice, for a given set of the physical parameters $\{\Omega
_{k},\Omega _{l},J_{kl}\}$ obtained according to the above theoretical
design, the Hamiltonians $H$ and $H_{e}$ are determined by Eqs.(2) and (4),
respectively. By choosing the evolutional time as $t_{p}=\left| (2m+1)\pi
/\Omega _{k}^{^{\prime }}\right| $ the propagator $\exp (-iH_{e}t_{p})$ will
take really either the form: $V^{2}\exp (-i\pi I_{kz})$ (Eq.(9)) or the
form: $V^{2}\exp (i\pi I_{kz})$. Although the latter form is different from
Eq.(9), the same result of the echo sequence $P_{1}$ of Eq.(11) is obtained.
On the other hand, the parameter $\lambda _{kl}$ in Eq.(12) will take really 
$n\pi +4\alpha _{1}^{+}$, where $\alpha _{1}^{+}=\arctan \{-\dfrac{\pi J_{kl}%
}{\Omega _{l}}-\dfrac{1}{2}\dfrac{\Omega _{k}}{\Omega _{l}}(-\delta +\sqrt{%
\delta ^{2}+4})\}$ is determined from Eq.(16b) by the above theoretical
design, but the factor $n\pi $ does not give a net effect on the two-qubit
diagonal gate of Eq.(12). Thus, the two-qubit diagonal gate $G_{kl}(\lambda
_{kl})$ $(\lambda _{kl}=4\alpha _{1}^{+})$ is really prepared exactly by the
echo sequence (12).

As an example, to prepare the two-qubit XOR gate $^{12}$ whose unitary
operation can be expressed as

$U_{XOR}=\exp (i\frac{\pi }{4})\exp (-i\frac{\pi }{2}I_{ly})\exp (-i\frac{%
\pi }{2}I_{kz})\exp (-i\frac{\pi }{2}I_{lz})$

$\qquad \qquad \times \exp (-i\pi I_{kz}I_{lz})\exp (i\frac{\pi }{2}I_{ly})$%
\newline
one needs to prepare the two-qubit diagonal gate with $\lambda _{kl}=\frac{%
\pi }{2}$. For $\lambda _{kl}=\frac{\pi }{2}$ and $\gamma ^{+}=\frac{3}{2}\
(m=1)$ the parameters $\mu $ and $\nu $ are obtained from Eqs.(17): $\mu =%
\frac{26-5\sqrt{2}}{24}$ and$\ \nu =\frac{5}{24}\sqrt{2}$ and $\Omega
_{k}^{^{\prime }}$ from Eq.(15a): $\frac{\Omega _{k}^{^{\prime }}}{\Omega
_{k}}=\pm \frac{1}{4}\sqrt{26-5\sqrt{2}}$, respectively. The evolutional
time is therefore determined by $\left| \Omega _{k}\right| t_{p}=\frac{12\pi 
}{\sqrt{26-5\sqrt{2}}}.$\newline
\newline
\textbf{Acknowledgement}

This work was supported by the NSFC general project with grant number:
19974064\newline
\newline
\textbf{References}\newline
\newline
1. R.P.Feynman, Int.J.Theor.Phys. 21, 467(1982)\newline
2. D.Deutsch, Proc.R.Soc.Lond. A 400, 97(1985)\newline
3. P.Shor, Proceedings of the 35th Annual Symposium on Foundations of

Computer Science, S.Goldwasser, ed., IEEE Computer Society,

Los Alamitos, CA, 1994, pp.124\newline
4. L.Grover, Phys.Rev.Lett. 79, 325(1997)\newline
5. D.Deutsch, Proc.R.Soc.Lond. A 425, 73(1989)\newline
6. S.Lloyd, Science 261, 1569(1993)\newline
7. D.P.DiVincenzo, Science 270, 255(1995)\newline
8. D.P.DiVincenzo, Phys.Rev. A 51, 1015(1995)\newline
9. T.Toffoli, Math.System Theory, 14, 13(1981)\newline
10. A.Barenco, Proc.R.Soc.Lond. A 449, 679(1995)\newline
11. T.Sleator and H.Weinfurter, Phys.Rev.Lett. 74, 4087(1995)\newline
12. A.Barenco, et al., Phys.Rev. A52, 3457(1995)\newline
13. X.Miao, http://xxx.lanl.gov/abs/quant-ph/0003068 (2000)\newline
14. J.I.Cirac and P.Zoller, Phys.Rev.Lett. 74, 4091(1995)\newline
15. Q.A.Turchette, C.J.Hood, W.Lange, H.Mabuchi, and H.J.Kimble,

Phys.Rev.Lett. 75, 4710(1995)\newline
16. N.A.Gershenfeld and I.L.Chuang, Science 275, 350(1997)\newline
17. D.G.Cory, A.F.Fahmy, and T.F.Havel, Proc.Natl.Acad.Sci.USA 94, 1634(1997)%
\newline
18. A.Shnirman, G.Sch$\ddot{o}$n, Z.Hermon, Phys.Rev.Lett. 79, 2371(1997)%
\newline
19. D.V.Averin, Solid State Commun. 105, 659(1998)\newline
20. Yu.Makhlin, G.Sch$\ddot{o}$n, A.Shnirman, Nature 398, 305(1999)\newline
21. Y.Nakamura, Yu.A.Pashkin, J.S.Tsai, Nature 398, 786(1999)\newline
22. P.Benioff, J.Stat.Phys. 22, 563(1980)\newline
23. X.Miao, Molec.Phys. 2000 (in press)\newline
24. R.R.Ernst, G.Bodenhausen, and A.Wokaun, \textit{Principles of Nuclear }

\textit{Magnetic Resonance in One and Two Dimensions.}

Oxford University Press, Oxford, 1987\newline
25. O.W.S$\phi $rensen, G.W.Eich, M.H.Levitt, G.Bodenhausen, R.R.Ernst,

Prog.NMR Spectrosc. 16, 163(1983)

\end{document}